\documentclass[aps,prl,reprint,superscriptaddress]{revtex4-2}
\usepackage{bm,amsmath,amssymb,color}
\usepackage{graphicx}

\begin{document}

\title{Diffusion Enhancement and Directional Suppression Induced by Reciprocal Flows}

\author{Yuki Koyano}
\email{koyano@garnet.kobe-u.ac.jp}
\affiliation{Graduate School of Human Development and Environment, Kobe University, Kobe 657-0011, Japan
}

\author{Hiroyuki Kitahata}
 \email{kitahata@chiba-u.jp}
\affiliation{Department of Physics, Graduate School of Science, Chiba University, Chiba 263-8522, Japan
}

\author{Takeshi Ooshida}
 \email{ooshida@tottori-u.ac.jp}
 \affiliation{Department of Mechanical and Physical Engineering, Tottori University, Tottori 680-8552, Japan}

\date{\today}

\begin{abstract}
Reciprocal flows repeatedly return fluid elements to their initial positions, producing no net advective transport on time average.
Nevertheless, their interplay with diffusion gives rise to nontrivial transport.
To describe this phenomenon, we develop a general theory of effective diffusion under two-dimensional linear flows with arbitrary time dependence.
By analyzing the advection-diffusion equation, we derive exact expressions for the mean square displacement and the effective diffusion coefficients for extensional, simple shear, and rotational flows within a single framework. 
We show that the reciprocal flows universally induce diffusion enhancement.
The diffusion tensor exhibits pronounced anisotropy, which can result in directional diffusion suppression in spite of the direction-averaged diffusion enhancement.
Our results provide a general framework for diffusion control by time-dependent flows and can provide new strategies for transport manipulation in microfluidic and biological systems.
\end{abstract}

\maketitle

The interplay between advection and diffusion can enhance the spreading of passive tracers, as exemplified by classical Taylor–Aris dispersion~\cite{Taylor_rspa.1953.0139, Taylor_rspa.1954.0216, Aris_rspa.1956.0065, Young1991, Frankel_JFM1989}. 
In some cases, such interplay is characterized by the {P\'eclet} number~\cite{Goldstein10.1098/rsfs.2015.0030},  $\mathrm{Pe} = v_0 L/D_0$,  which is the ratio of the externally posed length scale $L$ to the internal length scale $D_0/v_0$ determined by the representative flow velocity $v_0$ and the pure diffusion coefficient $D_0$ originating from thermal fluctuations.
There are other cases, however, in which diffusion is enhanced by flows without such a characteristic length scale $L$.
For example, effective diffusion coefficients have been studied for extensional flows~\cite{Mauri_SIAM} and simple shear flows~\cite{VANDEVEN1977,Katayama_1996,Orihara_2011,Orihara_2013,Orihara_2019,Mauri_SIAM, SANMIGUEL1979,Vilquin_PRF2021}.
These flows are typical examples of linear flows, i.e., flows whose velocity is a linear function of the position. 
In two-dimensional systems, they are classified into three types: extensional, simple shear, and rotational flows~\cite{MASON1977275}.
Diffusion enhancement by all three types of flows has also been studied~\cite{Bernard_SIAM,Goddard_PhysFluidsA,MIYAZAKI199553,Morris_Brady_1996,Compte_PhysRevE.55.6821,Foister_Ven_1980}.

While diffusion enhancement by steady flows has been extensively studied, studies on those by unsteady flows remain limited, in spite of the ubiquity of time-dependent flows in cytoplasmic streaming~\cite{Goldstein10.1098/rsfs.2015.0030},  protoplasmic streaming~\cite{Sato2025e220002}, and micro\-fluidic devices~\cite{LEE2014183}.
Diffusion enhancement by oscillatory flows has been studied in a confined pipe flow~\cite{Watson_1983,Aris_1960} and in a levitated droplet~\cite{Watanabe2018, Koyano_etal_2020}.
In the context of linear flows, sinusoidal time dependence has been discussed for simple shear flow~\cite{Rhines_Young_1983,Leighton1989,orihara2012} and for all three types of flow~\cite{Indeikina_PhysFluidsA}.
A remarkable exception is Krishnan \textit{et al.}~\cite{Krishnan1992}, who studied extensional flows with arbitrary time dependence.
Systematic studies on a broader class of unsteady flows are still required.
  
Motivated by this, here we present a systematic investigation of diffusion enhancement by unsteady linear flows in a two-dimensional setup.
As the study on flows with more complicated flow patterns \cite{Koyano_etal_2020} suggests that the presence of shear is important, it is a reasonable simplification to limit to linear flows.
We develop a general framework applicable to linear flows with effectively arbitrary time dependence.

\begin{figure}
    \centering
    \includegraphics{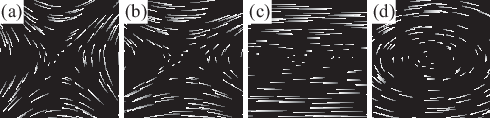}
    \caption{Typical examples of linear flows. Streamlines are shown for (a) the extensional flow at $a = b = 0.5$ (the principal axes are orthogonal), (b) the extensional flow at $a = 0.75$, $b = 0.25$, (c) the simple shear flow at $a = 1$, $b = 0$, and (d) the rotational flow at $a= 1.5$, $b = -0.5$.}
    \label{fig1}
\end{figure}

We start with an advection-diffusion equation for a two-dimensional concentration field $c(\bm{r},t)$:
\begin{align}
    \frac{\partial c}{\partial t} + \bm{v}\cdot \nabla c = D_0 \nabla^2 c, \label{RDeq}
\end{align}
where $D_0$ is the diffusion coefficient originating from thermal fluctuations. We introduce the Cartesian coordinates $\bm{r} = {}^t(x,y) = {}^t(x_1,x_2) $.
We assume an incompressible flow field $\bm{v}(\bm{r},t)$ in which the temporal and spatial dependencies are separated and the spatial part is a linear function, i.e.,
\begin{align}
    \bm{v}(\bm{r},t) = \dot{\gamma}(t)\mathsf{A} \bm{r} = \dot{\gamma}(t) 
    \begin{pmatrix} 
    0 & a \\
    b & 0
    \end{pmatrix}
    \begin{pmatrix}
    x \\ y
    \end{pmatrix}, \label{flowfield_general}
\end{align}
where $\dot{\gamma}(t)$ denotes the instantaneous magnitude of the flow with constant parameters $a$ and $b$ determining the shape of the streamlines. By the redefinition of the coordinate axes, we can obtain the form in Eq.~\eqref{flowfield_general} with $a \geq \left|b\right|$.
The generality of Eq.~\eqref{flowfield_general} is justified in ~Sec.~\ref{app_Z} of Supplemental Material (SM). The eigenvalues of $\mathsf{A}$ are $\pm (ab)^{1/2}$.
The spatial profile of the flow is classified by the sign of $ab$ into three types~\cite{MASON1977275}, namely extensional flow ($ab > 0$), simple shear flow ($a > b = 0$), and rotational flow ($ab < 0$). The three cases are exemplified in Fig.~\ref{fig1}.
It should be noted that $a = -b$ corresponds to a rigid-body-like rotation, which does not affect diffusion. Thus we exclude this case.

We consider the case where the flow does not cause net advection over a long-time average. We define a function corresponding to the strain $\gamma(t)$ as
\begin{align}
    \gamma(t) = \int_0^t \dot{\gamma}(t') dt', \label{F}
\end{align}
which is required to be a bounded continuous function.
A typical example of $\dot{\gamma}(t)$ is a sinusoidal function.

To discuss diffusion under unsteady flows, we consider the second moments of the concentration field $c(\bm{r},t)$:
\begin{align}
    M_{ij}(t) &= \frac{\iint x_i x_j c(\bm{r},t) \, d\bm{r}}{\iint c(\bm{r},t) \, d\bm{r}}.
\label{def_moment}
\end{align}
Here, $c(\bm{r},t)$ is assumed to be localized so that $M_{ij}(t)$ is finite.
Multiplying both sides of Eq.~\eqref{RDeq} by $x_i x_j$ and integrating over the whole space, we obtain three equations for the moments, which we write in a form of a single equation for a three-component vector 
$\bm{M}(t) = {}^t(M_{xx}(t), M_{xy}(t), M_{yy}(t))$: 
\begin{align}
    \frac{d \bm{M}}{dt} 
    = \dot{\gamma}(t) \mathsf{B} \bm{M} + 2D_0 \bm{C}, \label{eq_main} 
\end{align}
where
\begin{align}
    \bm{C} = \begin{pmatrix} 1 \\ 0 \\ 1 \end{pmatrix}, \quad
    \mathsf{B} =\begin{pmatrix} 0 & 2a & 0 \\ b & 0 & a \\ 0 & 2b & 0 \end{pmatrix}.
\end{align}
Detailed derivation of Eq.~\eqref{eq_main} is shown in Sec.~\ref{calc_moment_eq} of SM.
We formally write down the solution of the inhomogeneous linear differential equation~\eqref{eq_main} as
\begin{align}
    \bm{M}(t) =& \exp(\gamma(t)\mathsf{B}) \bm{M}_0 \nonumber \\
    &+ 2D_0 \left[ \int_{0}^{t} \exp \left( \gamma_\Delta(t,t') \mathsf{B} \right) dt' \right] \bm{C}, \label{sol_M}
\end{align}
using $\bm{M}_0 = \bm{M}(0)$ and $\gamma_\Delta(t,t') = \gamma(t) - \gamma(t')$ for brevity.
If the initial condition of $c(\bm{r},t)$ is axisymmetric, we have
\begin{align}
    \bm{M}_0 = M_0 \bm{C} \quad (M_0 \geq 0). \label{initial_condition}
\end{align}
 In particular, if the initial profile of $c(\bm{r},t)$ is given by the Dirac delta function, $M_0$ vanishes. Under the initial condition in Eq.~\eqref{initial_condition}, the mean square displacement (MSD) of $c(\bm{r},t)$ is given by $Q(t)= M_{xx}(t) + M_{yy}(t) = \bm{C} \cdot \bm{M}(t)$, that is 
\begin{align}
    Q(t) =& M_0 \bm{C} \cdot \exp(\gamma(t) \mathsf{B}) \bm{C} \nonumber \\
    &+ 2D_0 \bm{C} \cdot \left[ \int_{0}^{t} \exp \left( \gamma_\Delta(t,t') \mathsf{B} \right) dt' \right] \bm{C}. \label{msd_nondegenerate}
\end{align}
Detailed derivations of  Eqs.~\eqref{sol_M} and \eqref{msd_nondegenerate} are shown in Sec.~\ref{eq_momentSM} of SM.
Using the spectral decomposition of $\mathsf{B}$, we obtain
\begin{align}
  & Q(t)= 2M_0 + 4D_0t \nonumber \\
   &+ M_0 \frac{(a+b)^2}{2ab} \left\{\cosh \left(2 \sqrt{ab} \gamma(t)\right) - 1 \right \} \nonumber \\
     &+ D_0 \frac{(a+b)^2}{ab} \int_0^t \left \{ \cosh \left(2 \sqrt{ab} \gamma_\Delta(t,t')\right) -1 \right \} dt' \label{dif_posn}
\end{align}
in the case of $ab >0$, i.e., extensional flows, and 
\begin{align}
   &Q(t)= 2M_0 + 4D_0t \nonumber \\
   &+ M_0 \frac{(a+b)^2}{-2ab} \left\{1 - \cos \left(2 \sqrt{-ab} \gamma(t)\right)\right \} \nonumber \\
     &+ D_0 \frac{(a+b)^2}{-ab} \int_0^t \left \{1 - \cos \left(2 \sqrt{-ab} \gamma_\Delta(t,t')\right) \right \} dt' \label{dif_negn}
\end{align}
in the case of $ab <0$, i.e., rotational flows. 
In the case of $a > b = 0$, i.e., simple shear flows, it is more convenient to calculate the matrix exponential directly, 
which yields
\begin{align}
    Q(t) 
   =&2 M_0 + 4D_0 t + M_0 a^2 \left\{\gamma(t)\right\}^2 \nonumber \\
    &
    + 2 D_0 a^2 \int_0^t \left\{ \gamma_\Delta(t, t')\right\}^2dt'. \label{Vdegenerate}
\end{align}
Detailed calculation is shown in Sec.~\ref{app_B} of SM.
It should be noted that both descriptions in Eqs.~\eqref{dif_posn} and \eqref{dif_negn} converge to Eq.~\eqref{Vdegenerate} in the limit of $b \to 0$.

Now let us discuss how to extract the information of the effective diffusion coefficient from $Q(t)$.
At first glance it may seem easy to define it as 
\begin{align}
    \mathcal{D}(t)=& \frac{Q(t) - Q(0)}{4t}
    \label{mathcal.D}, 
\end{align}
  which we refer to as an apparent diffusion coefficient. 
We note, however, that $\mathcal{D}(t)$ can be unduly influenced by advection if $\gamma(t) \ne 0$.
It is therefore sensible to focus on zeros of the function $\gamma(t)$, denoting the $m$-th zero of $\gamma(t)$ with $\tau_m$, and regard $\mathcal{D}(\tau_m)$ as the effective diffusion coefficient.
Assuming the convergence of $\mathcal{D}(\tau_m)$ for $m \to \infty$, we define
the long-time effective diffusion coefficient $D$ by 
\begin{align}
    D 
    = \lim_{m \to \infty} \mathcal{D}(\tau_m) 
    = \lim_{m \to \infty} \frac{Q(\tau_m)}{4\tau_m}. \label{LT-effective-diffusion}
\end{align}
Note that if there exist constants $0 < T_{\mathrm{min}} < T_{\mathrm{max}}$ such that $0 < T_{\mathrm{min}} \leq \tau_{m+1}-\tau_m \leq T_{\mathrm{max}}$ for all $m$, $\mathcal{D}(\tau_{m+1})-\mathcal{D}(\tau_m) \to 0$ for $m\to \infty$ is proved, which is a necessary but not sufficient condition for convergence of the sequence $\{\mathcal{D}(\tau_m)\}$. See details in Sec.~\ref{app_limit} in SM.

Considering that $\gamma_\Delta(\tau_m,t') = - \gamma(t')$, we obtain
\begin{align}
    \frac{D}{D_0} = 1 + \lim_{m \to \infty} \frac{(a+b)^2}{4ab \tau_m} \int_0^{\tau_m} \left \{ \cosh \left(2 \sqrt{ab} \gamma(t')\right) -1 \right \} dt' \label{dif_pos}
\end{align}
for $ab > 0$,
\begin{align}
\frac{D}{D_0} = 1 + \lim_{m\to \infty} \frac{(a+b)^2}{-4ab\tau_m} \int_0^{\tau_m} \left \{1 - \cos \left(2 \sqrt{-ab} \gamma(t')\right) \right \} dt' \label{dif_neg}
\end{align}
for $ab<0$, and
\begin{align}
   \frac{D}{D_0} =1 + \lim_{m \to \infty} \frac{a^2}{2\tau_m} \int_0^{\tau_m} \left\{ \gamma(t')\right\}^2dt'
   \label{dif_0}
\end{align}
for $a > b = 0$.
In every case, the second term on the right-hand side is positive since the integrand is nonnegative for all $t'$, with a trivial exception $\gamma(t) \equiv 0$. 

Let us evaluate the effective diffusion concretely in the case of
\begin{align}
    \dot{\gamma}(t) = f_0 \cos(\omega t + \varphi) \label{sinusoidal}
\end{align}
implying basically $\tau_{m+2} - \tau_m =T$, with $T=2\pi/\omega$ denoting the period.
In this case, the long-time effective diffusion coefficient $D$ exists and is explicitly calculated as:
\begin{align}
    \frac{D}{D_0}=& 1 + \frac{(a+b)^2}{4 ab} \nonumber \\
    &\times \left[\cosh \left(\frac{2 f_0 \sqrt{ab} \sin\varphi}{\omega} \right) \mathcal{I}_0\left(\frac{2f_0\sqrt{ab}}{\omega}\right) - 1 \right] \label{dif_pos_cos}
\end{align}
for $ab > 0$,
\begin{align}
    \frac{D}{D_0} =& 1 + \frac{(a+b)^2}{-4 ab} \nonumber \\
    &\times\left[1 - \cos \left(\frac{2f_0\sqrt{-ab} \sin\varphi}{\omega} \right) \mathcal{J}_0\left(\frac{2f_0\sqrt{-ab}}{\omega}\right) \right], \label{dis_neg_cos}
\end{align}
for $ab < 0$, and
\begin{align}
     \frac{D}{D_0} = 1 + \frac{{f_0}^2 a^2}{4\omega^2}(1 + 2\sin^2 \varphi) \label{dif_0_cos}
\end{align}
for $a > b = 0$.
Here, $\mathcal{I}_0(\cdot)$ and $\mathcal{J}_0(\cdot)$ are modified and ordinary Bessel functions of the first kind of zeroth order, respectively. 
Note that all the second terms on the right-hand sides of Eqs.~\eqref{dif_pos_cos}--\eqref{dif_0_cos} are positive, indicating that a sinusoidally varying, spatially linear flow field always enhances diffusion. Since these terms are normalized by $D_0$ and explicitly depend on the flow field, the enhancement can be interpreted as a cooperative effect of advection and diffusion. Detailed calculations are shown in Sec.~\ref{app_D} of SM.
The results in Eqs.~\eqref{dif_pos_cos}--\eqref{dif_0_cos} as well as the second moments $\bm{M}(t)$, which are explicitly shown in Sec.~\ref{app_M} in SM, are consistent with the results in the previous papers~\cite{Rhines_Young_1983,Leighton1989,orihara2012,Indeikina_PhysFluidsA}.

In order to confirm these theoretical results, we perform numerical simulation of the advection-diffusion equation~\eqref{RDeq}. 
The flow profile and its time dependence are prescribed in Eq.~\eqref{flowfield_general} and Eq.~\eqref{sinusoidal}, respectively.
The calculation is based on the finite difference method for spatial discretization, combined with an explicit time-integration scheme; details are given in Sec.~\ref{app_D+} of SM. 
The initial condition is prescribed to be Gaussian, 
  $c(\bm{r}, 0) = \exp(-(x^2+y^2)/(2w^2))$ with small $w$. 
The time step and spatial mesh are set to be $\Delta t = T / (3\times 10^6)$  and $\Delta x = \ell/20$, where $\ell$ is a reference scale that can be chosen arbitrarily.
Here we choose $\ell$ so as to satisfy $D_0 t_\mathrm{max}/\ell^2 \sim 1$, allowing time evolution until $t_\mathrm{max} = 5T$ (i.e.~five oscillation periods).
We also impose the constraint $a+b=1$, as a kind of normalization of $\mathsf{A}$.

In what follows, all the quantities are nondimensionalized with $\omega$ and $\ell$.
Numerical constants are set to $D_0 = 0.01$ and $w = 0.1$; note that $\omega$ becomes unity by definition.
The flow amplitude $f_0$ and the initial phase $\varphi$ are fixed to 1 and 0, respectively, unless otherwise specified. 
The computational domain is taken to be $-10 \leq x \leq 10$ and $-10\leq y \leq 10$, and the Dirichlet boundary condition for $c$ is imposed. 
In the conditions adopted here, the flux almost vanishes at the boundary within our calculation time, so the boundary condition hardly affects the results.

\begin{figure}
    \centering
    \includegraphics{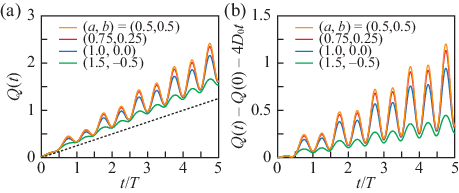}
    \caption{Numerical simulation results. (a) MSDs $Q(t)$ as functions of time $t$ for various cases of $(a,b)$. The dotted line corresponds to the MSD for the case without advective flows, i.e. $4D_0 t$. (b) Increment of MSDs from the pure diffusion $Q(t) - Q(0) - 4D_0t$. Here, $4D_0t$ is the pure diffusion originating from the thermal fluctuations and $Q(0)$ is the initial variance of $c$. $a = b = 0.5$, yellow; $a = 0.75$, $b = 0.25$, red; $a = 1.0$, $b = 0.0$, blue; $a = 1.5$, $b = -0.5$, green.}
    \label{fig2}
\end{figure}

The time series of the second moments, Eq.~\eqref{def_moment}, are computed from the numerical data of $c$. In Fig.~\ref{fig2}a, the MSDs $Q(t)$ are displayed for extensional, simple shear, and rotational flows. 
For all three flow types, MSDs oscillate at the half oscillation period of the flow velocity.  As the MSDs in the presence of flows always exceed that in their absence (cf. dotted line in Fig.~\ref{fig2}a), we extract this excess by subtracting the initial MSD, $Q(0)(= 2M_0)$ (i.e., the variance of $c(0)$), and the pure diffusive contribution, $4D_0t$, from the MSD, $Q(t)$, which is shown in Fig.~\ref{fig2}b. Evidently, the diffusion is enhanced, with the excess asymptotically proportional to $t$. This observation is consistent with the theoretical predictions given by Eqs.~\eqref{dif_pos_cos}--\eqref{dif_0_cos}.

To characterize the anisotropy of the diffusion process, we introduce the apparent diffusion tensor,
\begin{align}
    \mathcal{D}_{ij}(t) = \frac{M_{ij}(t) - M_{ij}(0)}{2t}. \label{effective_diffusion_tensor}
\end{align}
We refer to its eigenvalues, $\mathcal{D}_\pm(t)$ ($\mathcal{D}_+(t) \geq \mathcal{D}_-(t)$), as the apparent directional diffusion coefficients, since they represent the diffusion coefficients along the corresponding eigenvector directions.
For an isotropic initial profile of $c(0)$, i.e., $M_{ij}(0) = M_0 \delta_{ij}$, 
the average of the eigenvalues, $\mathcal{D}(t)= (\mathcal{D}_+ (t) + \mathcal{D}_- (t))/2$, is equal to the apparent diffusion coefficient in Eq.~\eqref{mathcal.D}. The details are given in Sec.~\ref{app_E} of SM.
Note that $\mathcal{D}_{ij} = D_0 \delta_{ij}$ for the case without hydrodynamic flow, where $\delta_{ij}$ is the Kronecker delta.

\begin{figure}
    \centering
    \includegraphics{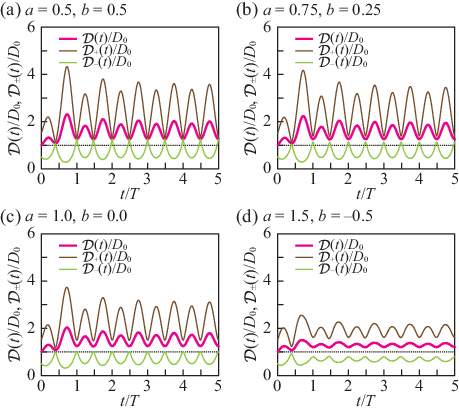}
    \caption{Normalized apparent diffusion coefficient $\mathcal{D}(t)/D_0$ (magenta thick curve) and its anisotropy $\mathcal{D}_{\pm}(t)/D_0$ ($\mathcal{D}_+(t)/D_0$: brown thin curve, $\mathcal{D}_-(t)/D_0$: light green thin curve) for each flow profile obtained by numerical simulations. (a) $a = b = 0.5$. (b) $a = 0.75$, $b = 0.25$. (c) $a = 1.0$, $b = 0.0$. (d) $a = 1.5$, $b = -0.5$. Black dotted lines indicate the case without any cooperative effect ($\mathcal{D}(t)/D_0 = \mathcal{D}_\pm(t)/D_0 = 1$).}
    \label{fig3}
\end{figure}

Time series of the apparent directional diffusion coefficients are shown in Fig.~\ref{fig3}, which exhibits an unexpected behavior: the apparent diffusion in the eigenvector direction corresponding to $\mathcal{D}_-(t)$ can be smaller than the pure diffusion $D_0$ in certain cases.
We plot the normalized apparent directional diffusion coefficients, $\mathcal{D}_{\pm}(t)/D_0$, for the three types of flow, together with the normalized apparent diffusion coefficient, $\mathcal{D}(t)/D_0$.
For extensional and simple shear flow, $\mathcal{D}_-(nT)$ ($n \in \mathbb{N}$) exceeds $D_0$ as shown in Fig.~\ref{fig3}a--c, while $\mathcal{D}_-(nT)$ falls below $D_0$ for rotational flow as in Fig.~\ref{fig3}d, i.e., apparent directional diffusion suppression occurs.
Such apparent directional diffusion suppression can be explained by the analytical calculation of $\mathcal{D}_-(t)$ (See Sec.~\ref{app_F} of SM.
We obtain $\mathcal{D}_-(\tau_m) < D_0$ for rotational flows ($ab < 0$) and $\mathcal{D}_-(\tau_m) \leq D_0$ for simple shear flows ($a > b = 0$), where the equality holds for $\int_0^{\tau_m} \gamma(t') dt' = 0$.
This means that apparent directional diffusion suppression always occurs for rotational flow and in most cases for simple shear flow. For extensional flow, apparent directional diffusion suppression may or may not occur, depending on the time dependence of the shear rate $\dot{\gamma}(t)$, and can therefore be controlled. For example, apparent directional diffusion suppression is predicted for all three types of flow by setting $\dot{\gamma}(t) = f_0 \cos (\omega t + \pi/2)$.  As shown in Sec.~\ref{app_phi_pi2} of the SM, the numerical results confirm that apparent directional diffusion suppression indeed occurs.

\begin{figure}
    \centering
    \includegraphics{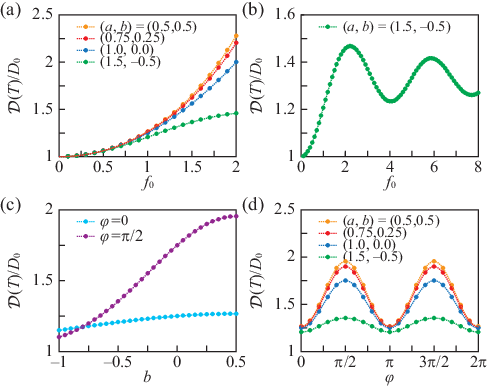}
    \caption{Parameter dependence of $\mathcal{D}(T)/D_0$. Points show the numerical simulation results, while the dotted curves show the theoretical results in Eqs.~\eqref{dif_pos_cos}--\eqref{dif_0_cos}. (a,b) $\mathcal{D}(T)/D_0$ versus $f_0$. The range of $f_0$ is broader in (b). (c) $\mathcal{D}(T)/D_0$ versus $b$. 
    (d) $\mathcal{D}(T)/ D_0$ versus the initial phase of flow $\varphi$. In panels (a), (b), and (d), plot colors correspond to Fig.~\ref{fig2}. In panel (c), cyan and purple curves indicate $\varphi = 0$ and $\pi/2$, respectively. 
    }
    \label{fig4}
\end{figure}

In Fig.~\ref{fig4}, we confirm that the theoretical results in Eqs.~\eqref{dif_pos_cos}--\eqref{dif_0_cos} quantitatively agree with the numerical simulation results.
We evaluate $\mathcal{D}(t)/D_0$ at $t = T$, and compare it with the theoretical results. 
In Fig.~\ref{fig4}a, the dependence of $\mathcal{D}(T)/D_0$ on the flow amplitude $f_0$ is shown. The diffusion enhancement becomes stronger with an increase in $f_0$ for extensional and simple shear flows. In the case of rotational flow, by contrast, $\mathcal{D}(T)/D_0$ oscillates for larger $f_0$ as shown in Fig.~\ref{fig4}b. 
The maxima and minima of $\mathcal{D}(T)/D_0$ are interpreted to arise when the tracer particles in the flow field rotate by approximately $(2n-1)\pi$ and $2n\pi$ ($n \in \mathbb{N}$) during a half period, $T/2$, respectively.
The value of $\mathcal{D}(T)/D_0$ also depends on the flow profile characterized by the parameter $b$ as shown in Fig.~\ref{fig4}c. The diffusion enhancement becomes stronger for extensional flows ($b>0$) than for rotational flows ($b<0$).
The dependence of $\mathcal{D}(T)/D_0$ on the initial phase $\varphi$ of the flow profile oscillation is also shown in Fig.~\ref{fig4}d.

In this paper, we have investigated the effective diffusion under a linear flow with effectively arbitrary time dependence. The MSDs and effective diffusion coefficients for each flow type are generally obtained as in Eqs.~\eqref{dif_posn}--\eqref{Vdegenerate} and Eqs.~\eqref{dif_pos}--\eqref{dif_0}, respectively.
There have been previous studies on effective diffusion under linear flows with sinusoidal dependence~\cite{Rhines_Young_1983,Leighton1989,Indeikina_PhysFluidsA,orihara2012} and also under extensional flows with arbitrary dependence~\cite{Krishnan1992}.  The present results provide a generalization of these works to all the three types of linear flows with arbitrary time dependence; the present formulas are applicable to reciprocal flows with effectively arbitrary time dependence, including nonperiodic and sufficiently regular stochastic time profiles as well as sinusoidal, square and triangular waves.
We also demonstrate directional diffusion suppression; $\mathcal{D}_-$ can fall below the equilibrium diffusion coefficient $D_0$. The diffusion can be suppressed in a certain direction by the advective flow, though the apparent diffusion coefficient averaged over all directions always exceeds the diffusion without flows.

Our results indicate the feasibility of diffusion control by flows, which can be beneficial for microfluidic technology.
From an application perspective, the efficiency of diffusion enhancement (or suppression) with respect to the dissipated energy remains an interesting topic for future work.

\begin{acknowledgments}
This work was supported by JSPS KAKENHI Grants Nos.~JP24K06978, JP24K22311, JP24K16981, JP25K00918, JP26K00671, MEXT KAKENHI Grant No.~JP26H00384, JSPS Bilateral Program Number JSPSBP120264601, and the Cooperative Research Program of ``Network Joint Research Center for Materials and Devices'' (Nos.~20254003, 20251011, 20261048, 20264003).
This work was also supported by the JSPS Core-to-Core Program ``Advanced core-to-core network for the physics of self-organizing active matter (JPJSCCA20230002)'' and by MEXT Promotion of Distinctive Joint Usage/Research Center Support Program Grant Number JPMXP0724020292.
\end{acknowledgments}

\providecommand{\noopsort}[1]{}\providecommand{\singleletter}[1]{#1}%

\clearpage

\newpage

\setcounter{page}{1}
\onecolumngrid

\begin{center}
\textbf{\large Supplemental Material: Diffusion Enhancement and Directional Suppression Induced by Reciprocal Flows}
\end{center}

\begin{center}
{\normalsize Yuki Koyano, Hiroyuki Kitahata, Takeshi Ooshida}
\end{center}
\setcounter{secnumdepth}{3}
\renewcommand{\theequation}{S\arabic{equation}}
\setcounter{equation}{0}

\renewcommand{\thefigure}{S\arabic{figure}}
\setcounter{figure}{0}

\thispagestyle{empty}
\section{Derivation of generic form of the flow field\label{app_Z}}

Flow fields $\tilde{\bm{v}}(\tilde{\bm{r}},t)$ with a uniform shear are generically written in the form
\begin{align}
    \tilde{\bm{v}}(\tilde{r},t) = \dot{\gamma}(t) \tilde{\mathsf{A}} \tilde{\bm{r}} = \dot{\gamma}(t) \begin{pmatrix} \tilde{a}_{xx} & \tilde{a}_{xy} \\ \tilde{a}_{yx} & \tilde{a}_{yy}  \end{pmatrix} \tilde{\bm{r}}. 
\end{align}
From the incompressibility condition,
\begin{align}
\nabla \cdot \tilde{\bm{v}} =\dot{\gamma}(t) \left( \tilde{a}_{xx} + \tilde{a}_{yy} \right) = 0 
\end{align}
holds. 

If $\tilde{a}_{xy} \geq \tilde{a}_{yx}$, we consider the rotational transform of the coordinates $\bm{r}$ as
\begin{align}
    \bm{r} = \begin{pmatrix} \cos \Theta & \sin \Theta \\ - \sin\Theta & \cos\Theta  \end{pmatrix} \tilde{\bm{r}},
\end{align}
with
\begin{align}
    \cos 2\Theta = \frac{\tilde{a}_{xy} + \tilde{a}_{yx}}{\sqrt{4 {\tilde{a}_{xx}}^2 + \left(\tilde{a}_{xy} + \tilde{a}_{yx}\right)^2}},
\qquad
    \sin 2\Theta = -\frac{2\tilde{a}_{xx} }{\sqrt{4 {\tilde{a}_{xx}}^2 + \left(\tilde{a}_{xy} + \tilde{a}_{yx}\right)^2}}.
\end{align}
Then, we get
\begin{align}
    \bm{v}(\bm{r},t) = \dot{\gamma}(t) \begin{pmatrix} 0 & \tilde{a} \\ \tilde{b} & 0 \end{pmatrix} \bm{r},
\end{align}
with
\begin{align}
    a = \frac{\tilde{a}_{xy} - \tilde{a}_{yx}}{2} + \frac{\sqrt{4 {\tilde{a}_{xx}}^2 + \left(\tilde{a}_{xy} + \tilde{a}_{yx}\right)^2}}{2} , \qquad
    b = - \frac{\tilde{a}_{xy} - \tilde{a}_{yx}}{2} + \frac{\sqrt{4 {\tilde{a}_{xx}}^2 + \left(\tilde{a}_{xy} + \tilde{a}_{yx}\right)^2}}{2}.
\end{align} 

If $\tilde{a}_{xy} < \tilde{a}_{yx}$, we consider the rotational transform and mirror reflection of the coordinates $\bm{r}$ as
\begin{align}
    \bm{r} = \begin{pmatrix} 0 & 1 \\ 1 & 0 \end{pmatrix} \begin{pmatrix} \cos \Theta & \sin \Theta \\ - \sin\Theta & \cos\Theta  \end{pmatrix} \tilde{\bm{r}},
\end{align}
with
\begin{align}
    \cos 2\Theta = \frac{\tilde{a}_{xy} + \tilde{a}_{yx}}{\sqrt{4 {\tilde{a}_{xx}}^2 + \left(\tilde{a}_{xy} + \tilde{a}_{yx}\right)^2}},
\qquad
    \sin 2\Theta = \frac{2\tilde{a}_{xx} }{\sqrt{4 {\tilde{a}_{xx}}^2 + \left(\tilde{a}_{xy} + \tilde{a}_{yx}\right)^2}}.
\end{align}
Then, we get
\begin{align}
    \bm{v}(\bm{r},t) = \dot{\tilde{\gamma}}(t) \begin{pmatrix} 0 & \tilde{a} \\ \tilde{b} & 0 \end{pmatrix} \bm{r},
\end{align}
with
\begin{align}
    a = \frac{\tilde{a}_{yx} - \tilde{a}_{xy}}{2} + \frac{\sqrt{4 {\tilde{a}_{xx}}^2 + \left(\tilde{a}_{xy} + \tilde{a}_{yx}\right)^2}}{2} ,
\qquad
    b = - \frac{\tilde{a}_{yx} - \tilde{a}_{xy}}{2} + \frac{\sqrt{4 {\tilde{a}_{xx}}^2 + \left(\tilde{a}_{xy} + \tilde{a}_{yx}\right)^2}}{2}.
\end{align} 
It is noted that  $a \geq \left|b\right|$ is satisfied by considering the cases according to the relative magnitudes of $\tilde{a}_{xy}$ and $\tilde{a}_{yx}$.

Therefore, we only need to consider the case in Eq.~\eqref{flowfield_general} in the main text.

\section{Derivation of the time evolution equation of the moments in Eq.~(\ref{eq_main}) in the main text \label{calc_moment_eq}}

Since $c(\bm{r},t)$ is localized, we get
\begin{align}
    \iint x^m y^n \frac{\partial c}{\partial x} dx \, dy 
    =& \int \left\{ \left[x^m c\right]_{x \to -\infty}^{x \to +\infty} - \int m x^{m-1} c \, dx  \right\} y^n dy \nonumber \\
    =& -m \iint x^{m-1}y^n c \, dx \,dy,  
    \label{intx}
\end{align}
for nonnegative integers $m$ and $n$ and 
\begin{align}
    \iint x^m y^n \frac{\partial c}{\partial y} dx \, dy 
    = -n \iint x^my^{n-1}c \, dx \,dy  ,\label{inty}
\end{align}
in a parallel manner.
We also get
\begin{align}
    \iint x^m y^n \frac{\partial^2 c}{\partial x^2} dx \, dy 
    =& \int \left\{ \left[x^m \frac{\partial c}{\partial x} \right]_{x \to -\infty}^{x \to +\infty} - \int m x^{m-1} \frac{\partial c}{\partial x} \, dx  \right\} y^n dy \nonumber \\ 
    =& -m \iint x^{m-1}y^n \frac{\partial c}{\partial x}  dx \, dy  \nonumber  \\
     =& m (m-1) \iint x^{m-2}y^n c \, dx \, dy, \label{intxx}
\end{align}
and 
\begin{align}
     \iint x^m y^n \frac{\partial^2 c}{\partial y^2} dx \, dy
     = n (n-1) \iint x^m y^{n-2}c \, dx \, dy, \label{intyy}
\end{align}
in a parallel manner. 

Equation \eqref{RDeq} in the main text is divided by $\int c(\bm{r},t)d\bm{r}$ for normalization. Then,
by multiplying $x^2$ and integrating over the whole space, we obtain
\begin{align}
    \frac{dM_{xx}}{dt} - 2a \dot{\gamma}(t) M_{xy} = 2 D_0.
\end{align}
In the same manner, by multiplying $xy$ and $y^2$, we obtain
\begin{align}
    \frac{dM_{xy}}{dt} -b \dot{\gamma}(t)  M_{xx} - a \dot{\gamma}(t)  M_{yy} = 0, 
\end{align}
and
\begin{align}
    \frac{dM_{yy}}{dt} -2b \dot{\gamma}(t)  M_{xy} = 2 D_0, 
\end{align}
respectively.
Here, we use Eq.~\eqref{flowfield_general} in the main text and Eqs.~\eqref{intx}--\eqref{intyy}. Thus, we obtain Eq.~\eqref{eq_main} in the main text.

\section{Detailed calculation to obtain Eqs.~(\ref{sol_M}) and (\ref{msd_nondegenerate}) in the main text \label{eq_momentSM}}

Considering that Eq.~\eqref{eq_main} in the main text is an inhomogeneous linear differential equation, the solution can be written in the form:
\begin{align}
    \bm{M}(t) = \exp \left(\gamma(t) \mathsf{B} \right) \bm{Y}(t), \label{M}
\end{align}
where $\gamma(t)$ is defined in Eq.~\eqref{F} in the main text.

By substituting Eq.~\eqref{M} into Eq.~\eqref{eq_main} in the main text, we have the time evolution equation for $\bm{Y}(t)$ as
\begin{align}
    \frac{d\bm{Y}}{dt} = 2D_0 \exp \left(- \gamma(t) \mathsf{B}\right) \bm{C}. \label{eq_Y}
\end{align}
Although $\mathsf{B}$ is singular, the matrix exponential $\exp(\gamma(t)\mathsf{B})$ is always invertible, with inverse $\exp(-\gamma(t)\mathsf{B})$.
%Note that even in the case with $ab = 0$, i.e., $\mathsf{B}$ does not have an inverse matrix, $\exp \left(\gamma(t) \mathsf{B}\right)$ has an inverse matrix $\exp \left(-\gamma(t) \mathsf{B}\right)$.

From Eq.~\eqref{initial_condition} in the main text, the initial condition of $\bm{Y}(t)$ is given by
\begin{align}
    \bm{Y}(0) = \bm{M}(0) = \bm{M}_0. 
\end{align}
Thus, the solution of Eq.~\eqref{eq_Y} is expressed as
\begin{align}
    \bm{Y}(t) = \bm{M}_0 + 2D_0 \left [ \int_{0}^{t} \exp \left( - \gamma(t') \mathsf{B}\right) dt' \right] \bm{C}.
\end{align}
Thus, we have the solution
\begin{align}
    \bm{M}(t) =& \exp \left(\gamma(t) \mathsf{B} \right) \bm{M}_0 + 2D_0 \left[ \int_{0}^{t} \exp \left\{ \left( \gamma(t) - \gamma(t') \right) \mathsf{B} \right\} dt' \right] \bm{C} \nonumber \\
    =& \exp \left(\gamma(t) \mathsf{B} \right) \bm{M}_0 + 2D_0 \left[ \int_{0}^{t} \exp \left( \gamma_\Delta(t, t') \mathsf{B} \right) dt' \right] \bm{C}.
\end{align}
Thus, Eq.~\eqref{sol_M} in the main text is derived.

Using the result above, the mean square displacement (MSD) $Q(t)= M_{xx}(t) + M_{yy}(t) = \bm{C} \cdot \bm{M}(t)$ in the case where $\bm{M}_0 = M_0 \bm{C}$ is obtained as
\begin{align}
    Q(t) =& M_0 \bm{C} \cdot \exp \left(\gamma(t)\mathsf{B}\right) \bm{C} + 2D_0 \bm{C} \cdot \left[ \int_{0}^{t} \exp \left( \gamma_\Delta(t, t') \mathsf{B} \right)  dt' \right] \bm{C},
\end{align}
which appears in Eq.~\eqref{msd_nondegenerate} in the main text.

\section{Detailed calculation for Eqs.~(\ref{dif_posn})--(\ref{Vdegenerate}) in the main text and explicit expression for $\bm{M}(t)$ \label{app_B}}

In this section, we obtain the simplified forms of Eqs.~\eqref{sol_M} and \eqref{msd_nondegenerate} in the main text.
The eigenvalues of the matrix $\mathsf{B}$ are given as $0$ and $\pm 2(ab)^{1/2}$. 

In the case where $ab \neq 0$, the spectral decomposition of $\mathsf{B}$ is introduced as
\begin{align}
    \mathsf{B} = \sum_{i} \lambda^{(i)} \mathsf{B}^{(i)},
\end{align}
where $\lambda^{(1)} = 0$, $\lambda^{(2)} = 2(ab)^{1/2}$, $\lambda^{(3)} = -2 (ab)^{1/2}$, and $\mathsf{B}^{(i)}$ is a projection matrix with respect to the eigenvalue $\lambda^{(i)}$. They satisfy the following relation:
\begin{align}
\mathsf{B}^{(i)} \mathsf{B}^{(j)} = \delta_{ij} \mathsf{B}^{(i)}, \label{orth_rel}  
\end{align}
where $\delta_{ij}$ is the Kronecker delta.
Considering Eq.~\eqref{orth_rel}, we have
\begin{align}
    \exp\left(\xi \mathsf{B}\right) =
        \exp\left(\xi \sum_{i=1}^3 \lambda^{(i)}  \mathsf{B}^{(i)} \right) 
        = \sum_{i=1}^3 \exp\left(\lambda^{(i)} \xi\right) \mathsf{B}^{(i)}
\end{align}
for any scalar variable $\xi$.

Therefore, Eq.~\eqref{msd_nondegenerate} in the main text becomes
\begin{align}
    Q(t)  
    &= M_0 \sum_{i=1}^3 \exp (\lambda^{(i)} \gamma(t)) \bm{C} \cdot \mathsf{B}^{(i)} \bm{C} + 2D_0 \int_{0}^{t} \sum_{i=1}^3 \left [  \exp \left \{ \lambda^{(i)} \gamma_\Delta(t,t') \right \} \bm{C} \cdot \mathsf{B}^{(i)} \bm{C} \right ] dt'. \label{case_pos}
\end{align}
The matrices $\mathsf{B}^{(i)}$ for $i = 1,2,3$ are explicitly given as
\begin{align}
    \mathsf{B}^{(1)} = \frac{1}{2} \begin{pmatrix}
        1 & 0 & -a/b \\
        0 & 0 & 0 \\
        -b/a & 0 & 1
    \end{pmatrix},
\end{align}
\begin{align}
    \mathsf{B}^{(2)} = \frac{1}{4} \begin{pmatrix}
        1 & 2(a/b)^{1/2} & a/b   \\
        (b/a)^{1/2} & 2 & (a/b)^{1/2} \\
        b/a & 2(b/a)^{1/2} & 1
    \end{pmatrix},
\end{align}
\begin{align}
    \mathsf{B}^{(3)} = \frac{1}{4} \begin{pmatrix}
        1 & - 2(a /b)^{1/2} & a/b   \\
        - (b/a)^{1/2} & 2 &  -(a/b)^{1/2} \\
        b/a & - 2( b/a)^{1/2} & 1
    \end{pmatrix},
\end{align}
and thus
\begin{align}
    \bm{C}\cdot \mathsf{B}^{(1)} \bm{C} = \frac{(a-b)^2}{2ab},
\end{align}
\begin{align}
    \bm{C}\cdot \mathsf{B}^{(2)} \bm{C} = \frac{(a+b)^2}{4ab},
\end{align}
\begin{align}
    \bm{C}\cdot \mathsf{B}^{(3)} \bm{C} = \frac{(a+b)^2}{4ab}.
\end{align}
Using these results, we obtain
\begin{align}
    Q(t)=& M_0 \left[ -\frac{(a-b)^2}{2ab} + \frac{(a+b)^2}{2 ab} \cosh \left(2 \sqrt{ab} \gamma(t) \right)\right]+ 2D_0\int_0^t \left[ -\frac{(a-b)^2}{2ab} + \frac{(a+b)^2}{2ab} \cosh \left(2 \sqrt{ab} \gamma_\Delta(t,t')\right)\right] dt' \nonumber \\
     =& 2M_0 + M_0 \frac{(a +b)^2}{2 ab} \left\{\cosh (2 \sqrt{ab} \gamma(t)) - 1 \right\} + 4D_0t + D_0 \frac{(a+b)^2}{ab}  \int_0^t \left\{\cosh \left(2 \sqrt{ab} \gamma_\Delta(t,t')\right) -1 \right\} dt'
\end{align}
for $ab > 0$, and
\begin{align}
    Q(t)=& M_0 \left[ \frac{(a-b)^2}{-2ab} - \frac{(a+b)^2}{-2 ab} \cos \left(2 \sqrt{-ab} \gamma(t) \right)\right]+ 2D_0\int_0^t \left[\frac{(a-b)^2}{-2ab} - \frac{(a+b)^2}{-2ab} \cos \left(2 \sqrt{-ab} \gamma_\Delta(t,t')\right)\right] dt' \nonumber \\
     =& 2M_0 + M_0 \frac{(a +b)^2}{-2 ab} \left\{1 - \cos (2 \sqrt{-ab} \gamma(t))\right\} + 4D_0t + D_0 \frac{(a+b)^2}{-ab}  \int_0^t \left\{1 - \cos \left(2 \sqrt{-ab} \gamma_\Delta(t,t')\right)  \right\} dt'
\end{align}
for $ab<0$.

In the case where $ab=0$, i.e., $b = 0$, $\mathsf{B}^3 = \mathsf{O}$ holds, and thus we have
\begin{align}
\exp\left( \xi \mathsf{B} \right) = \mathsf{E} + \xi \mathsf{B} + \frac{1}{2}\xi^2 \mathsf{B}^2 = \begin{pmatrix}
1 & 2a \xi & a^2 \xi^2 \\
0 & 1 & a \xi \\
0 & 0 & 1
\end{pmatrix}.
\end{align}
Here, $\mathsf{O}$ and $\mathsf{E}$ are zero and unit matrices, respectively. Then, we obtain
\begin{align}
    Q(t) =& M_0 \left[2 + a^2 \left\{\gamma(t)\right\}^2 \right] + 2D_0 \int_0^t \left[2 + a^2 \left\{ \gamma_\Delta(t,t')\right\}^2 \right] dt' \nonumber \\
    =& 2M_0 + M_0 a^2 \left\{\gamma(t)\right\}^2 + 4D_0t + 2D_0 a^2 \int_0^t \left\{\gamma_\Delta(t,t')\right\}^2 dt'.
\end{align}

\section{Asymptotic behavior of the effective diffusion coefficient \label{app_limit}}

Here, we consider the asymptotic behavior of the effective diffusion
coefficient evaluated at $\tau_m$, defined by
$\gamma(\tau_m)=0$.
The expressions in Eqs.~\eqref{dif_pos}--\eqref{dif_0} of the main text can be written in the generic form
\begin{align}
d_m = \frac{D(\tau_m)}{D_0}
=
1 + \frac{A}{\tau_m}
\int_0^{\tau_m}
P\left(\gamma(t')\right)\,dt',
\label{eq:dm_general}
\end{align}
where $A$ is a prefactor and $P(\gamma)$ is a continuous nonnegative function.
For example, $P(\gamma)$ corresponds to
$\cosh(2\sqrt{ab}\gamma)-1$ for extensional flows,
$1-\cos(2\sqrt{-ab}\gamma)$ for rotational flows,
and $\gamma^2$ for simple shear flows.

%\begin{equation}
%0<T_{\min} \le \tau_{m+1}-\tau_m \le T_{\max}.
%\label{eq:tau_interval}
%\end{equation}

Using Eq.~\eqref{eq:dm_general}, $d_{m+1}$ can be rewritten as
\begin{align}
d_{m+1} &= 1+ \frac{A}{\tau_{m+1}} \int_0^{\tau_{m+1}}
P\left(\gamma(t')\right)\,dt'
\nonumber\\
&=
1+ \frac{A}{\tau_{m+1}}
\int_0^{\tau_m} P\left(\gamma(t')\right)\,dt'
+
\frac{A}{\tau_{m+1}}
\int_{\tau_m}^{\tau_{m+1}}
P\left(\gamma(t')\right)\,dt'
\nonumber\\
&=
1+ \frac{\tau_m}{\tau_{m+1}}
(d_m-1) + \frac{A}{\tau_{m+1}}
\int_{\tau_m}^{\tau_{m+1}}
P\left(\gamma(t')\right)\,dt'.
\label{eq:dm_recurrence}
\end{align}
Therefore,
\begin{align}
d_{m+1}-d_m =
-\frac{\tau_{m+1}-\tau_m}{\tau_{m+1}}
(d_m-1) + \frac{A}{\tau_{m+1}}
\int_{\tau_m}^{\tau_{m+1}}
P\left(\gamma(t')\right)\,dt'.
\label{eq:dm_difference}
\end{align}

Since $\gamma(t)$ is bounded, we set $\gamma_{\min}
\le \gamma(t) \le \gamma_{\max}$. From the continuity of $P(\gamma)$, $0 \le P\left(\gamma(t)\right) \le P_{\max}$ holds. 
Considering that $0<T_{\min} \le \tau_{m+1}-\tau_m \le T_{\max}$, 
$\tau_{m+1}\ge \tau_0 + (m+1)T_{\min}$ holds. Using these conditions, we obtain 
\begin{align}
\left|d_{m+1}-d_m\right|
&\le
\frac{T_{\max}}{\tau_0 + (m+1)T_{\min}}
\left|d_m-1\right|
+
\frac{|A|P_{\max}T_{\max}}
{\tau_0 + (m+1)T_{\min}}.
\label{eq:dm_difference_bound}
\end{align}
Furthermore, Eq.~(\ref{eq:dm_general}) gives
\begin{equation}
|d_m-1|
\le
|A|P_{\max},
\end{equation}
and hence the right-hand side of
Eq.~(\ref{eq:dm_difference_bound}) vanishes as $m\to\infty$.
Consequently,
\begin{equation}
\lim_{m\to\infty}
\left(d_{m+1}-d_m\right)=0.
\label{eq:successive_difference}
\end{equation}

Thus, the difference between the effective diffusion coefficients
evaluated at $\tau_m$ and $\tau_{m+1}$ vanishes in the long-time limit.  We note that Eq.~(\ref{eq:successive_difference}) alone does
not guarantee the convergence of the sequence $\{d_m\}$. Therefore, in the main
text, the long-time effective diffusion coefficient $D$ is defined
only when the limit $\lim_{m\to\infty}D(\tau_m)$ exists.

\section{Detailed calculation in the case of sinusoidal reciprocal flow \label{app_D}}

Here, we calculate Eqs.~\eqref{dif_pos}--\eqref{dif_0} in the main text in the case of $\dot{\gamma} = f_0 \cos (\omega t + \varphi)$.
In this case, $\gamma(t) = 0$ holds at $t = 2 n \pi / \omega = nT$ with $n \in \mathbb{N}$, where $T$ is the period. Thus, we calculate the effective diffusion coefficient at $t = 2n \pi / \omega = nT$.

If $ab > 0$, the effective diffusion coefficient in Eq.~\eqref{dif_pos} in the main text is calculated as
\begin{align}
    \frac{D}{D_0} =& 1 + \lim_{n \to \infty} \frac{(a+b)^2}{4abnT} \int_0^{nT} \left \{ \cosh \left(2 \sqrt{ab} \gamma(t')\right) -1 \right \} dt' \nonumber \\
    =& 1 + \lim_{n \to \infty} \frac{(a+b)^2}{8ab nT} \int_0^{nT} \left[\exp \left(  \frac{2 f_0\sqrt{ab}}{\omega} \left( \sin(\omega t' + \varphi) - \sin \varphi \right) \right)  \right. \nonumber \\
     & \qquad \qquad \qquad \qquad \qquad \qquad \left.+ \exp \left( -\frac{2 f_0\sqrt{ab}}{\omega} \left( \sin(\omega t' + \varphi) - \sin \varphi \right) \right)- 2\right] dt' \nonumber \\
     =& 1 + \lim_{n \to \infty} \frac{(a+b)^2}{8ab nT} nT \left[   \exp\left( -\frac{2f_0\sqrt{ab} \sin\varphi}{\omega}\right) \mathcal{I}_0\left(\frac{2f_0\sqrt{ab}}{\omega} \right) + \exp\left(  \frac{2f_0\sqrt{ab} \sin\varphi}{\omega}\right)  \mathcal{I}_0\left(\frac{2f_0\sqrt{ab}}{\omega}\right) - 2 \right] \nonumber \\
    = & 1 + \frac{(a+b)^2}{4ab} \left[\cosh \left(\frac{2f_0\sqrt{ab} \sin\varphi}{\omega} \right) \mathcal{I}_0\left(\frac{2f_0\sqrt{ab}}{\omega}\right) -1\right].
\end{align}
If $ab < 0$, the effective diffusion coefficient in Eq.~\eqref{dif_neg} in the main text is calculated as
\begin{align}
    \frac{D}{D_0} =& 1 + \lim_{n \to \infty} \frac{(a+b)^2}{-4ab nT} \int_0^{nT} \left \{1 - \cos \left(2 \sqrt{-ab} \gamma(t')\right)\right \} dt' \nonumber \\
    =& 1 + \lim_{n \to \infty} \frac{(a+b)^2}{-8ab nT} \int_0^{nT} \left[2-\exp \left( i \frac{2 f_0\sqrt{-ab}}{\omega} \left( \sin(\omega t' + \varphi) - \sin \varphi \right) \right) \right. \nonumber \\
     & \qquad \qquad \qquad \qquad \qquad \qquad \left. + \exp \left( -i\frac{2 f_0\sqrt{-ab}}{\omega} \left( \sin(\omega t' + \varphi) - \sin \varphi \right) \right)\right] dt' \nonumber \\
     =& 1 + \lim_{n \to \infty} \frac{(a+b)^2}{-8ab nT} nT \left[   2 - \exp\left( -i\frac{2f_0\sqrt{-ab} \sin\varphi}{\omega}\right) \mathcal{J}_0\left(\frac{2f_0\sqrt{-ab}}{\omega} \right) \right. \nonumber \\
     & \qquad \qquad \qquad \qquad \qquad \qquad \left.- \exp\left(  i\frac{2f_0\sqrt{-ab} \sin\varphi}{\omega}\right)  \mathcal{J}_0\left(\frac{2f_0\sqrt{-ab}}{\omega}\right) \right] \nonumber \\
    = & 1 + \frac{(a+b)^2}{-4ab} \left[1 - \cos \left(\frac{2f_0\sqrt{-ab} \sin\varphi}{\omega} \right) \mathcal{J}_0\left(\frac{2f_0\sqrt{-ab}}{\omega}\right) \right].
\end{align}
If $a > b = 0$, the effective diffusion coefficient in Eq.~\eqref{dif_0} in the main text is calculated as
\begin{align}
    \frac{D}{D_0} =& 1 + \lim_{n\to \infty} \frac{a^2}{2nT} \int_0^{nT} \left(\frac{f_0}{\omega}\right)^2\left\{ \sin(\omega t' + \varphi) - \sin \varphi \right\}^2 dt' \nonumber\\
    =& 1 + \lim_{n\to \infty} \frac{a^2}{2nT} \left(\frac{f_0}{\omega}\right)^2 nT  \left( \frac{1}{2}+ \sin^2\varphi \right) \nonumber \\
    =& 1 + \frac{{f_0}^2a^2}{4\omega^2} \left(1 + 2 \sin^2 \varphi\right).
\end{align}

\section{Expression of the second moment $\bm{M}(t)$ \label{app_M}}

Using Eq.~\eqref{sol_M} in the main text, the vector $\bm{M}(t)$, composed of second moments, is calculated as
\begin{align}
    \bm{M}(t) =& \bm{M}_0 + \frac{\cosh\left(2\sqrt{ab}\gamma(t)\right) - 1}{2ab} \begin{pmatrix}
    ab & 0 & a^2 \\
    0 & 2ab & 0 \\
    b^2 & 0 & ab
    \end{pmatrix}\bm{M}_0 + \frac{\sinh\left(2\sqrt{ab}\gamma(t)\right)}{2\sqrt{ab}} \begin{pmatrix}
    0 & 2 a & 0 \\
    b & 0 & a \\
    0 & 2b & 0
    \end{pmatrix}\bm{M}_0 \nonumber \\
   &+ 2D_0t \bm{C} + D_0 \frac{a+b}{ab} \int_0^t 
   \begin{pmatrix}
   a \left\{ \cosh \left(2\sqrt{ab} \gamma_\Delta(t,t')\right) - 1 \right\} \\
    \sqrt{ab} \sinh \left(2\sqrt{ab} \gamma_\Delta(t,t')\right) \\
   b \left\{ \cosh \left(2\sqrt{ab} \gamma_\Delta(t,t')\right) - 1 \right\}    
   \end{pmatrix}  dt' \label{Mtm_pos}
\end{align}
for $ab>0$, 
\begin{align}
    \bm{M}(t) =& \bm{M}_0 + \frac{1 - \cos\left(2\sqrt{-ab}\gamma(t)\right) }{-2ab} \begin{pmatrix}
    ab & 0 & a^2 \\
    0 & 2ab & 0 \\
    b^2 & 0 & ab
    \end{pmatrix}\bm{M}_0 + \frac{\sin\left(2\sqrt{-ab}\gamma(t)\right)}{2\sqrt{-ab}} \begin{pmatrix}
    0 & 2 a & 0 \\
    b & 0 & a \\
    0 & 2b & 0
    \end{pmatrix}\bm{M}_0 \nonumber \\
   &+ 2D_0t \bm{C} + D_0  \frac{a+b}{-ab} \int_0^t 
   \begin{pmatrix}
   a \left\{ 1 - \cos \left(2\sqrt{-ab} \gamma_\Delta(t,t')\right) \right\} \\
    \sqrt{-ab} \sin \left(2\sqrt{-ab} \gamma_\Delta(t,t')\right) \\
   b \left\{1 - \cos \left(2\sqrt{-ab} \gamma_\Delta(t,t')\right) \right\}    
   \end{pmatrix}  dt' \label{Mtm_neg}
\end{align}
for $ab<0$, and
\begin{align}
    \bm{M}(t) =& \bm{M}_0 + \left\{\gamma(t)\right\}^2 \begin{pmatrix}
    0 & 0 & a^2 \\
    0 & 0 & 0 \\
    0 & 0 & 0
    \end{pmatrix}\bm{M}_0 + \gamma(t) \begin{pmatrix}
    0 & 2 a & 0 \\
    0 & 0 & a \\
    0 & 0 & 0
    \end{pmatrix}\bm{M}_0+ 2D_0t \bm{C} + D_0  \int_0^t 
   \begin{pmatrix}
   2a^2 \left\{ \gamma_\Delta(t,t')\right\}^2 \\
    2a \gamma_\Delta(t,t') \\ 0    
   \end{pmatrix}  dt' \label{Mtm_0}
\end{align}
for $a > b = 0$.

The components of $\bm{M}(t)$ at $t = \tau_m$, where $\gamma(\tau_m) = 0$ holds, are given as
\begin{align}
    \bm{M}(\tau_m) =& \bm{M}_0+ 2D_0 \tau_m \bm{C} + D_0 \frac{a+b}{ab} \int_0^{\tau_m} 
   \begin{pmatrix}
   a \left\{ \cosh \left(2\sqrt{ab} \gamma(t')\right) - 1 \right\} \\
    -\sqrt{ab} \sinh \left(2\sqrt{ab} \gamma(t')\right) \\
   b \left\{ \cosh \left(2\sqrt{ab} \gamma(t')\right) - 1 \right\}    
   \end{pmatrix}  dt' \label{M_pos}
\end{align}
for $ab>0$, 
\begin{align}
    \bm{M}(\tau_m) =& \bm{M}_0 + 2D_0 \tau_m \bm{C} + D_0  \frac{a+b}{-ab} \int_0^{\tau_m} 
   \begin{pmatrix}
   a \left\{ 1 - \cos \left(2\sqrt{-ab} \gamma(t')\right) \right\} \\
    -\sqrt{-ab} \sin \left(2\sqrt{-ab} \gamma(t')\right) \\
   b \left\{1 - \cos \left(2\sqrt{-ab} \gamma(t')\right) \right\}    
   \end{pmatrix}  dt' \label{M_neg}
\end{align}
for $ab<0$, and
\begin{align}
    \bm{M}(\tau_m) =& \bm{M}_0 + 2D_0 \tau_m \bm{C} + 2D_0  \int_0^{\tau_m} 
   \begin{pmatrix}
   a^2 \left\{ \gamma(t')\right\}^2 \\
    -a \gamma(t') \\ 0    
   \end{pmatrix}  dt' \label{M_zero}
\end{align}
for $a > b = 0$.

Using Eqs.~\eqref{M_pos}--\eqref{M_zero}, $\bm{M}(t)$ at $t = nT$ ($n \in \mathbb{N}$) under the assumption of $\bm{M}_0 = \bm{0}$ is given as
\begin{align}
    \bm{M}(nT) = 2D_0 nT \bm{C} + D_0 nT \frac{a+b}{ab} \begin{pmatrix} a \left[ \cosh \left(\frac{2f_0\sqrt{ab} \sin\varphi}{\omega} \right) \mathcal{I}_0\left(\frac{2f_0\sqrt{ab}}{\omega}\right) -1 \right] \\ -\sqrt{ab} \sinh \left(\frac{2f_0\sqrt{ab} \sin\varphi}{\omega}\right) \mathcal{I}_0\left(\frac{2f_0\sqrt{ab}}{\omega}\right)  \\ b  \left[ \cosh \left(\frac{2f_0\sqrt{ab} \sin\varphi}{\omega} \right) \mathcal{I}_0\left(\frac{2f_0\sqrt{ab}}{\omega}\right) -1 \right] \end{pmatrix}
\end{align}
for $ab > 0$,
\begin{align}
    \bm{M}(nT) = 2D_0 nT \bm{C} + D_0 nT \frac{a+b}{-ab} \begin{pmatrix} a \left[1- \cos \left(\frac{2f_0\sqrt{-ab} \sin\varphi}{\omega}\right) \mathcal{J}_0\left(\frac{2f_0\sqrt{-ab}}{\omega}\right) \right] \\ -\sqrt{-ab} \sin \left(\frac{2f_0\sqrt{-ab} \sin\varphi}{\omega}\right) \mathcal{J}_0\left(\frac{2f_0\sqrt{-ab}}{\omega}\right)  \\ b  \left[ 1- \cos \left(\frac{2f_0\sqrt{-ab} \sin\varphi}{\omega} \right) \mathcal{J}_0\left(\frac{2f_0\sqrt{-ab}}{\omega}\right) \right] \end{pmatrix}
\end{align}
for $ab < 0$, and
\begin{align}
    \bm{M}(nT) = 2D_0 nT \bm{C} + D_0 nT \begin{pmatrix} \left(\frac{f_0 a}{\omega} \right)^2 \left(1 + 2\sin^2 \varphi \right) \\ -2 \frac{f_0 a}{\omega} \sin \varphi \\ 0 \end{pmatrix}
\end{align}
for $a > b = 0$.

\section{Numerical method in detail \label{app_D+}}

Here, we describe detailed information on the numerical simulation method.
The advection-diffusion equation~\eqref{RDeq} in the main text is discretized on a square lattice with spacing
$\Delta x=\Delta y$. The concentration field $c(\bm{r},t)$ is discretized as $c_{i,j}^n$, which denotes the value of the concentration field
$c(x,y,t)$ at the grid point $(x_i,y_j)$ and time $t_n$.
The lattice points are
\begin{equation}
  x_i=(i-i_0)\Delta x,\qquad
  y_j=(j-j_0)\Delta x,
\end{equation}
where $i = i_0$, $j = j_0$ correspond to the origin. 
The discretized time is 
\begin{align}
    t_n = n \Delta t,
\end{align}
where $\Delta t$ is chosen as $\Delta t= T/N_t$.
Here $T$ is the period of the reciprocal flow and $N_t$ is the number of time steps per oscillation period.

For calculating the time evolution of $c_{i,j}^n$ from $n$ to $n + 1$, we adopt
\begin{align}
    c_{i,j}^{n+1} = c_{i,j}^{n} + \Delta c_{i,j}^{n, \mathrm{adv}} + \Delta c_{i,j}^{n, \mathrm{dif}} 
\end{align}
The advection term is discretized in conservative form as
\begin{align}
  \Delta c_{i,j}^{n,\mathrm{adv}}
  =&
 - \Delta t \frac{
  v_x(x_{i+1/2},y_j,t_{n+1/2}) (c_{i+1,j}^n+c_{i,j}^n)
  -
  v_x(x_{i-1/2},y_j,t_{n+1/2}) (c_{i,j}^n+c_{i-1,j}^n)
  }{2\Delta x}
  \nonumber\\
  & -
  \Delta t \frac{
  v_y(x_i,y_{j+1/2},t_{n+1/2})(c_{i,j+1}^n+c_{i,j}^n)
  -
  v_y(x_i,y_{j-1/2},t_{n+1/2})(c_{i,j}^n+c_{i,j-1}^n)
  }{2\Delta x}.
\end{align}
Here, we set $\bm{v}(\bm{r},t) = {}^t(v_x(x,y,t), v_y(x,y,t))$ and adopt $\nabla \cdot (\bm{v} c) = \bm{v} \cdot \nabla c$ for incompressible flow fields $\bm{v}(\bm{r},t)$.
The diffusion term $\Delta c_{i,j}^{n,\mathrm{dif}}$ is evaluated by the standard second-order central
difference:
\begin{equation}
 \Delta c_{i,j}^{n,\mathrm{dif}}
  =
  D_0 \Delta t \frac{
  c_{i+1,j}^n+c_{i-1,j}^n+c_{i,j+1}^n+c_{i,j-1}^n-4c_{i,j}^n
  }{{\Delta x}^2}.
\end{equation}

The initial condition is a Gaussian distribution,
\begin{equation}
  c^0_{i,j}
  =
  \exp\left[-\frac{{x_i}^2+{y_j}^2}{2w^2}\right],
\end{equation}
where $w$ is the initial width. 
The calculation region is $-L \leq x \leq L$ and $-L \leq y \leq L$.
In the simulations shown in the main
text, $L=10$, $\Delta x=0.05$, $w = 0.1$, and
$N_t=3\times 10^6$.

Using the obtained results of $c_{i,j}^n$, the mean square displacement is evaluated as
\begin{align}
  Q(t_n)
  =
  \frac{\sum_{i,j}({x_i}^2+{y_j}^2) c^n_{i,j}}
       {\sum_{i,j} c^n_{i,j}},
\end{align}
which is used for discussion.

\section{Apparent diffusion tensor\label{app_E}}

The eigenvalues $\mathcal{D}_\pm(t)$ of the apparent diffusion tensor $\mathcal{D}_{ij}(t)$ defined in Eq.~\eqref{effective_diffusion_tensor} are explicitly given as
\begin{align}
\mathcal{D}_\pm(t) = \frac{\mathcal{D}_{xx}(t)+ \mathcal{D}_{yy}(t)}{2} \pm \frac{\sqrt{(\mathcal{D}_{xx}(t) - \mathcal{D}_{yy}(t))^2 + 4 {\mathcal{D}_{xy}(t)}^2}}{2}.
\end{align}
The sum of the eigenvalues is calculated
as
\begin{align}
    \mathcal{D}_+(t) + \mathcal{D}_-(t) = \mathcal{D}_{xx}(t) + \mathcal{D}_{yy}(t) = \frac{M_{xx}(t) + M_{yy}(t) - M_{xx}(0) - M_{yy}(0)}{2t} = \frac{Q(t) - Q(0)}{2t}.
\end{align}
Thus, 
\begin{align}
\mathcal{D}(t) = \frac{Q(t)-Q(0)}{4t} = \frac{\mathcal{D}_+(t) + \mathcal{D}_-(t)}{2} 
\end{align}
holds.

Considering that $M_{ij}(t)$ and $\mathcal{D}_{ij}(t)$ are symmetric tensors, i.e., $M_{ij}(t) = M_{ji}(t)$ and $\mathcal{D}_{ij}(t) = \mathcal{D}_{ji}(t)$, the eigenvectors are orthogonal to each other. The directions of the eigenvectors are regarded as principal directions, and $\mathcal{D}_+(t)$ and $\mathcal{D}_-(t)$ are the components of the apparent diffusion tensor along the principal directions.

\section{Explicit form of apparent diffusion tensor\label{app_F}}

The apparent diffusion tensor $\mathcal{D}_{ij}(t)$ is explicitly given as
\begin{align}
    \mathcal{D}_{ij}(t) = D_0 \delta_{ij}  +  \frac{\mathcal{M}_{ij}(t)}{2t},
\end{align}
where $\mathcal{M}_{ij}(t)$ is defined as
\begin{align}
    \mathcal{M}_{ij}(t) = M_{ij}(t) - M_{ij}(0) - 2 D_0 t \delta_{ij} 
\end{align}
Thus, the eigenvalues $\mathcal{D}_\pm(t)$ of $\mathcal{D}_{ij}(t)$ are represented using the eigenvalues $\mathcal{M}_\pm(t)$ of $\mathcal{M}_{ij}(t)$ as
\begin{align}
    \mathcal{D}_\pm(t) = D_0 + \frac{\mathcal{M}_\pm(t)}{2t}.
\end{align}

This indicates that the signs of $\mathcal{M}_\pm(\tau_m)$ determine whether apparent diffusion is enhanced or suppressed in their eigenvector directions. The trace $\mathcal{M}_{xx}(\tau_m) + \mathcal{M}_{yy}(\tau_m)$ of the tensor $\mathcal{M}_{ij}(\tau_m)$ is always positive from Eqs.~\eqref{M_pos}--\eqref{M_zero}. This shows that the effective diffusion is always enhanced.
In contrast, the sign of the determinant $\Delta_\mathcal{M}(\tau_m) = \mathcal{M}_{xx}(\tau_m) \mathcal{M}_{yy}(\tau_m) - {\mathcal{M}_{xy}(\tau_m)}^2$ is essential for the anisotropic diffusion enhancement or suppression; $\mathcal{M}_+(\tau_m) \geq \mathcal{M}_-(\tau_m) \geq 0$, i.e., $\mathcal{D}_+(\tau_m) - D_0 \geq \mathcal{D}_-(\tau_m) - D_0 \geq 0$, holds if $\Delta_\mathcal{M}(\tau_m) \geq 0$, while $\mathcal{M}_+(\tau_m) > 0 > \mathcal{M}_-(\tau_m)$, i.e., $\mathcal{D}_+(\tau_m) - D_0 > 0 > \mathcal{D}_-(\tau_m) - D_0$ holds if $\Delta_\mathcal{M}(\tau_m) < 0$.

From Eq.~\eqref{M_pos}, for extensional flows ($ab > 0$), we obtain
\begin{align}
    \Delta_\mathcal{M}(\tau_m) =& \frac{{D_0}^2(a+b)^2}{ab} \left( \left[ \int_0^{\tau_m} \left\{ \cosh \left(2 \sqrt{ab} \gamma(t')\right)-1\right\} dt' \right]^2 - \left[ \int_0^{\tau_m} \sinh \left(2 \sqrt{ab} \gamma(t')\right) dt' \right]^2 \right) \nonumber \\
    =& \frac{{D_0}^2(a+b)^2}{ab} \left[ \int_0^{\tau_m} \left\{ \exp \left(2 \sqrt{ab} \gamma(t')\right)-1\right\} dt' \right] \left[ \int_0^{\tau_m} \left\{ \exp \left(-2 \sqrt{ab} \gamma(t')\right) - 1\right\} dt' \right] .
\end{align}
Thus, the signs of the two components in the last equation determine whether diffusion is enhanced or suppressed in a certain direction. 

From Eq.~\eqref{M_neg}, for rotational flows ($ab < 0$), we obtain
\begin{align}
    \Delta_\mathcal{M}(\tau_m) = \frac{{D_0}^2(a+b)^2}{ab} \left( \left[ \int_0^{\tau_m} \left\{ 1 -\cos \left(2 \sqrt{-ab} \gamma(t')\right)\right\} dt' \right]^2 + \left[ \int_0^{\tau_m} \sin \left(2 \sqrt{-ab} \gamma(t')\right) dt' \right]^2 \right).
\end{align}
Thus, $\Delta_\mathcal{M}(\tau_m) \leq 0$ holds considering $ab<0$, and the equality holds only if $\gamma(t) \equiv 0$, which corresponds to the case without any convective flow.

From Eq.~\eqref{M_zero}, for simple shear flows ($a > b = 0$), we obtain
\begin{align}
    \Delta_M(\tau_m) = -4{D_0}^2a^2  \left( \int_0^{\tau_m}  \gamma(t') dt'\right)^2.
\end{align}
Therefore, $\Delta_\mathcal{M}(\tau_m) \leq 0$ holds. If $\int_0^{\tau_m} \gamma(t') dt' = 0$, $\Delta_M(\tau_m)$ is zero, which means no directional apparent diffusion suppression. 
It is notable that $\Delta_M(\tau_m) = 0$ means that the effective diffusion enhancement appears only in one direction, and no diffusion modification appears in the other direction.

Here, we explicitly calculate $\Delta_M(nT)$ in case that $\gamma(t)$ is sinusoidal as in Eq.~\eqref{sinusoidal}. 
For extensional flows ($ab>0$), $\Delta_M(nT)$ is calculated as
\begin{align}
    \Delta_M(nT) = \frac{{D_0}^2(a+b)^2}{ab} \left( nT\right)^2 \left[ \left\{\mathcal{I}_0\left(\frac{2f_0\sqrt{ab}}{\omega}\right) \right\}^2 - 2 \cosh\left(\frac{2 f_0 \sqrt{ab} \sin\varphi}{\omega} \right) \mathcal{I}_0\left(\frac{2f_0\sqrt{ab}}{\omega}\right) + 1 \right].
\end{align}
If $\varphi = 0$, then $\Delta_M(nT)$ is always positive. By expanding $\Delta_M(nT)$ with respect to $f_0\sqrt{ab}/\omega$, we obtain
\begin{align}
    \Delta_M(nT) =& \frac{{D_0}^2(a+b)^2}{ab} \left( nT \right)^2 \left[ -\left(\frac{2f_0\sqrt{ab}}{\omega}\right)^2 \sin^2\varphi + \frac{1}{48}\left(\frac{2f_0\sqrt{ab}}{\omega}\right)^4 \left(3 -12 \sin^2\varphi - 4\sin^4\varphi\right) \right] \nonumber \\
    &+ \mathcal{O}\left( \left(\frac{f_0 \sqrt{ab}}{\omega}\right)^6\right).
\end{align}
Therefore, $\Delta_M(nT)$ can be negative depending on $\varphi$.

Next, for rotational flows ($ab < 0$), we obtain
\begin{align*}
    \Delta_M(nT ) =& \frac{{D_0}^2(a+b)^2}{ab} \left( nT\right)^2 \left[ \left\{\mathcal{J}_0\left(\frac{2f_0\sqrt{-ab}}{\omega}\right) \right\}^2 - 2 \cos\left(\frac{2 f_0 \sqrt{-ab} \sin\varphi}{\omega} \right) \mathcal{J}_0\left(\frac{2f_0\sqrt{-ab}}{\omega}\right) + 1 \right].
\end{align*}
By considering that
\begin{align}
    &\left\{\mathcal{J}_0\left(\frac{2f_0\sqrt{-ab}}{\omega}\right) \right\}^2 - 2 \cos\left(\frac{2 f_0 \sqrt{-ab} \sin\varphi}{\omega} \right) \mathcal{J}_0\left(\frac{2f_0\sqrt{-ab}}{\omega}\right) + 1 \nonumber \\
    &= \left| \mathcal{J}_0\left(\frac{2f_0\sqrt{-ab}}{\omega}\right) \exp\left(\frac{2i f_0\sqrt{-ab} \sin\varphi}{\omega} \right)- 1 \right|^2 \geq 0,
\end{align}
we can claim that $\Delta_M(nT) \leq 0$. 

Finally, for simple shear flows ($a > b = 0$), we obtain
\begin{align}
\Delta_M(nT) = -4{D_0}^2 a^2 \left( \frac{f_0}{\omega} \right)^2 \left(nT\right)^2  \sin^2 \varphi.
\end{align}

\section{apparent directional diffusion coefficient for $\varphi = \pi/2$ \label{app_phi_pi2}}

\begin{figure}
    \centering
    \includegraphics{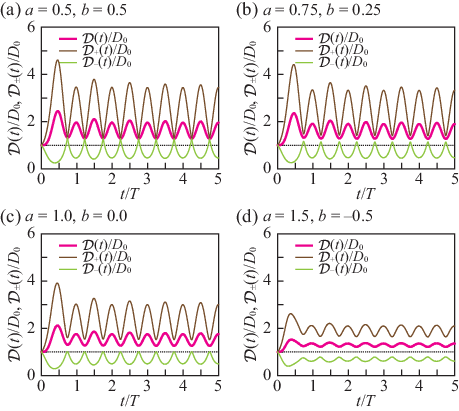}
     \caption{Numerical results corresponding to Fig.~\ref{fig3} in the main text with $\varphi = \pi/2$ instead of $\varphi = 0$. Normalized apparent diffusion coefficient $\mathcal{D}(t)/D_0$ (magenta thick curve) and normalized apparent directional diffusion coefficients $\mathcal{D}_\pm (t)/D_0$ ($\mathcal{D}_+(t)/D_0$: brown thin curve, $\mathcal{D}_-(t)/D_0$: light green thin curve) are shown. (a) $a = b = 0.5$. (b) $a = 0.75$, $b = 0.25$. (c) $a = 1.0$, $b = 0.0$. (d) $a = 1.5$, $b = -0.5$. Black dotted lines indicate the case without any cooperative effect ($\mathcal{D}(t)/D_0 = \mathcal{D}_\pm(t)/D_0 = 1$).}
    \label{fig_SM}
\end{figure}

The results with $\varphi = \pi/2$ instead of $\varphi = 0$ that correspond to Fig.~\ref{fig3} in the main text are shown in Fig.~\ref{fig_SM}. All of the values of $\mathcal{D}_-(nT)/D_0$ ($n \in \mathbb{N}$) are smaller than unity, which indicates the directional diffusion suppression.

\end{document}